\documentclass[12pt]{article}
\usepackage{amsbsy,latexsym}

\def\a{\alpha}

\def\C{\raise2pt\hbox{\rm\large$\chi$}}
\def\c{\chi}

\def\e{\epsilon}
\def\f{\phi}

\def\h{\eta}

\def\j{\psi}

\def\m{\mu}
\def\n{\nu}
\def\o{\omega}
\def\p{\pi}
\def\q{\theta}

\def\s{\sigma}

\def\z{\zeta}
\def\D{\Delta}

\def\vf{\varphi}

\def\bo{\bar{\o}}

\def\tf{\tilde{\f}}

\DeclareFontFamily{OT1}{msb}{}{}
\DeclareFontShape{OT1}{msb}{m}{n}
 {  <5> <6> <7> <8> <9> <10> gen * msbm
      <10.95><12><14.4><17.28><20.74><24.88>msbm10}{}
\DeclareMathAlphabet{\bubble}{OT1}{msb}{m}{n}

\def\bR{{\bubble R}}
\def\bZ{{\bubble Z}}

\def\ca{{\cal A}}

\def\cl{{\cal L}}
\def\cm{{\cal M}}

\def\cp{{\cal P}}

\def\cs{{\cal S}}

\def\det{{\rm det}}
\def\PCO{\cp}
\def\SFO{\cs}
\def\AZI{\ca}
\def\tPCO{\tilde{\cp}}
\def\hPCO{\hat{\cp}}

\def\pa{\partial}                                       

\def\bo{{\raise-.5ex\hbox{\large$\Box$}}}

\def\tr{{\rm tr}}

\def\ket#1{\left| #1\right\rangle}              
\def\VEV#1{\left\langle #1\right\rangle}        
\def\leftrightarrowfill{$\mathsurround=0pt \mathord\leftarrow \mkern-6mu
        \cleaders\hbox{$\mkern-2mu \mathord- \mkern-2mu$}\hfill
        \mkern-6mu \mathord\rightarrow$}
\def\dvec#1{\vbox{\ialign{##\crcr
        \leftrightarrowfill\crcr\noalign{\kern-1pt\nointerlineskip}
        $\hfil\displaystyle{#1}\hfil$\crcr}}}           
\def\dtt#1{{\buildrel \bullet \over {#1}}}              
\def\ad{\dtt{a}}
\def\bd{\dtt{b}}
\def\pd{\dtt{+}}
\def\md{\dtt{-}}

\def\fr#1#2{{\textstyle{#1\over\vphantom2\smash{\raise.20ex
        \hbox{$\scriptstyle{#2}$}}}}}                   
\def\partder#1#2{{\partial #1\over\partial #2}}   

\def\beq{\begin{equation}}
\def\eeq{\end{equation}}
\def\beqx{\begin{displaymath}} 
\def\eeqx{\end{displaymath}}
\def\beql{\arraycolsep .1em \begin{eqnarray}}
\def\eeql{\end{eqnarray}}
\def\zeile{\nonumber\\[.5ex] }
\def\gl#1{(\ref{#1})}

\def\theequation{\thesection.\arabic{equation}}

\catcode`@=11
\@addtoreset{equation}{section}
\@addtoreset{equation}{subsection}
\def\theequation{\ifnum\value{section}=0 \arabic{equation}\ignorespaces
\else \ifnum\value{section}=-1 A.\arabic{equation}\ignorespaces
\else \ifnum\value{subsection}=0 \thesection.\arabic{equation}\ignorespaces
\else \thesection.\arabic{subsection}.\arabic{equation}\ignorespaces
                           \fi
                      \fi
                 \fi}
\catcode`@=12

\parskip=\medskipamount

\begin{document}


\begin{titlepage}
\noindent
April, 1997 \hfill CERN--TH/97-59 \\
\phantom{X} \hfill ITP--UH--11/97 \\
hep-th/9704076 \hfill ITP--SB--97-20 

\vskip 1.0cm

\begin{center}

{\Large\bf N=2 WORLDSHEET INSTANTONS YIELD}

\medskip

{\Large\bf CUBIC SELF-DUAL YANG-MILLS~$^*$}

\vskip 1.5cm

{\large Olaf Lechtenfeld~$^+$}
 
{\it Theoretical Physics Division, CERN, CH--1211 Geneva 23}\\
{http://www.itp.uni-hannover.de/\~{}lechtenf/}\\

\vskip 0.7cm

{\large Warren Siegel}

{\it Institute for Theoretical Physics}\\
{\it State University of New York, Stony Brook, NY 11794-3840}\\
{siegel@insti.physics.sunysb.edu}\\

\vskip 1.5cm
\textwidth 6truein
{\bf Abstract}
\end{center}

\begin{quote}
\hspace{\parindent}
{}\ \ \
When the gauge instantons on the $N{=}2$ string worldsheet are properly
included in the sum over topologies, the breaking of $SO(2,2)$ Lorentz
symmetry in $\bR^{2,2}$ is parametrized by a {\it spacetime twistor\/}
containing the string coupling and theta angle. The resulting (tree-level)
effective action for the open string is not Yang's but Leznov's {\it cubic\/}
action for self-dual Yang-Mills in a light-cone gauge. In the closed case,
Pleba\'nski's action for self-dual gravity gets modified analogously.
In contrast to the $N{=}1$ NSR string, picture-changing is not locally
invertible, but produces a semi-infinite tower of massless physical states
with ever-increasing spin, perhaps related to harmonic superspace.
A truncation yields the two-field action of Chalmers and Siegel.
\end{quote}

\vfill

\textwidth 6.5truein
\hrule width 5.cm
\vskip.1in

{\small \noindent ${}^{*\ {}}$
supported in part by `Deutsche Forschungsgemeinschaft'
and `Volkswagen-Stiftung'}
\vskip-.1in
{\small \noindent {}\quad
as well as the National Science Foundation under grant no. PHY 9309888.\\
${}^+$ permanent address: 
Institut f\"ur Theoretische Physik, Universit\"at Hannover, Germany }

\eject
\end{titlepage}
\newpage
\hfuzz=10pt


Generally, the field or wave function of any bosonic string has only a
single component.  Usually, this implies that the ground state is a scalar.
However, the $N{=}2$ string describes self-dual Yang-Mills theory (open) or
self-dual gravity (closed)~\cite{OVold,marcus}.  
The description of either theory by a single component requires a unitary 
light-cone gauge, where the single polarization transforms nonlinearly under 
Lorentz transformations.  There are two well-known types of light-cone gauges 
for self-dual theories, which are ``dual'' to each other in the sense of 
trading a constraint with the field equation. 
For self-dual Yang-Mills theory, the Yang gauge~\cite{yang} results in a 
nonlinear field equation resembling that of a two-dimensional Wess-Zumino 
model, while the Leznov gauge~\cite{leznov,parkes} gives a quadratic field
equation.  The description of self-dual Yang-Mills theory originally found
from the $N{=}2$ string was in the Yang gauge~\cite{yang}.  Furthermore, 
in the closed string case the analog of the Yang gauge for gravity, the 
Pleba\'nski gauge~\cite{ple}, was found. However, there was no reason in 
principle why the string should prefer these gauges.  More recently, a gauge 
other than the Pleba\'nski gauge was found in the closed string case by 
including the previously ignored worldsheet $U(1)$ instantons~\cite{BL1}.
In this paper we identify that gauge as the gravity analog~\cite{siegel} of 
the Leznov gauge, and find also the Leznov gauge for the open string case.  
Furthermore, the two string couplings, associated with loop and instanton
number, are identified as the two-component commuting spinor (twistor) that 
chooses the arbitrary self-dual lightlike plane with respect to which the 
gauge is defined.

Strings with two world-sheet supersymmetries in the NSR formulation
are built from an $N{=}2$ world-sheet supergravity multiplet containing
the metric $h_{mn}$, an abelian gauge field $A_m$ and two charged Majorana
gravitini $\c^\pm_m$. 
In a self-dual $(2,2)$ metric background~$\cm$~\footnote{
Metric with $(r,r)$ signature have been termed `Kleinian'~\cite{klein}.
For $r{=}2$,
Kleinian self-duality implies Ricci flatness and $SL(2,\bR)$ holonomy.
Such spaces are called half-flat or hypersymplectic and possess one
complex and two real structures, all covariantly constant~\cite{klein}.}
the matter sector consists of the string coordinates $X$ 
and their charged NSR partners $\j$.
Specializing to flat space, $\cm=\bR^{2,2}$, we write
\beq
X^{a\ad}\ =\ \s_\mu^{a\ad} X^\mu\ =\
\left(\begin{array}{cc}
X^0{+}X^3 & X^1{+}X^2 \\ X^1{-}X^2 & X^0{-}X^3
\end{array}\right) \quad,\qquad 
a\in\{+,-\} \quad \ad\in\{\pd,\md\} \;,
\eeq
with a set of chiral gamma matrices $\s_\mu$, $\mu=0,\ldots,3$, 
appropriate for a spacetime metric $\h_{\mu\nu}={\rm diag}(-+-+)$.
Note that we are employing the van der Waerden index notation,
splitting $SO(2,2)$ vector indices $\mu$ into two $SL(2,\bR)$
spinor indices, $a$ and~$\ad$.
Spinor indices are raised and lowered by the epsilon tensor, and 
vectors have an $SL(2,\bR)\times~SL(2,\bR)'$ invariant length-squared of
\beq
\h_{\mu\nu}\ X^\mu X^\nu\ 
=\ -\fr12\,\e_{ab}\,\e_{\ad\bd}\ X^{a\ad}\,X^{b\bd}\
=\ -\det\,X^{a\ad} \quad.
\eeq

For later use, we define the following bilinears constructed from two
vectors $k$ and~$p$,
\beql\label{bilin}
p^a \wedge k^b \ &=&\ \e_{\ad\bd}\ p^{a\ad}\,k^{b\bd}
\zeile
p^{(+} \wedge k^{-)} \ &=&\ p^+\wedge k^- + p^-\wedge k^+
\zeile
p^{[+} \wedge k^{-]} \ &=&\ p^+\wedge k^- - p^-\wedge k^+ 
\ =\ \h_{\mu\nu}\ p^\mu k^\nu \quad,
\eeql
with contracted $SL(2,\bR)'$ indices suppressed.
We have diagonalized $L_{+-}$ as one of the 
$SL(2,\bR)$ {\it boost\/} generators, 
while the rotation is off-diagonal and recovered from the nilpotent 
light-cone combinations $L_{++}$ and $L_{--}$.
The three antisymmetric bilinears in \gl{bilin},
$p^+\wedge k^+$, $p^-\wedge k^-$, and $p^{(+}\wedge k^{-)}$, 
are still invariant under a smaller ``Lorentz group'' 
$\cl=GL(1,\bR)\times SL(2,\bR)'$, 
where the $GL(1,\bR)$ factor is created by 
$L_{++}$, $L_{--}$, or $L_{+-}$, respectively.

We are considering $n$-point amplitudes of oriented open strings.
In writing down the Brink-Schwarz action~\cite{brink}, 
we must pick some real or complex structure in $\bR^{2,2}$, 
with respect to which the action is neutral.
Consequently, only $\cl$ invariance is manifest.  
After superconformal gauge fixing,
string theory demands to average over matter fields $(X,\j)$ and the
ghost sector, to integrate over (metric, susy and gauge) moduli,
and to sum over worldsheet topologies parametrized by the genus
$g\in\frac12\bZ_+$ and the gauge instanton number $c\in\bZ$.
As is customary, contributions from different topologies are
weighted by powers of the string coupling~$e$ and the theta phase~$e^{i\q}$.
However, since we are working in a {\it real\/} basis, 
the latter combine to monomials in $\sin\q$ and $\cos\q$. 
The full string amplitude reads~\cite{berk1,OVnew,BL1}
\beql\label{amp}
A^{(n)}(k_1,\ldots,k_n)\ &=&\ \sum_{g,c}\ 
\Bigl(\begin{array}{c}\scriptstyle 2j\\ \textstyle{}^{j+c}\end{array}\Bigr)\
e^j\ \sin^{j-c}\fr{\q}{2} \cos^{j+c}\fr{\q}{2}\ \int\! dm_h\ dm_A
\zeile &\times&\ 
\VEV{\ V(k_1) \ldots V(k_n) \  \AZI\  \PCO_+^{j-c}\ \PCO_-^{j+c}\ }_{g,c}(m)
\eeql
where $j\equiv2g-2+n$, and a few remarks are in order.
\begin{itemize}
\item
Chan-Paton factors are implied but suppressed.
\item
For $|c|>j$, unbalanced susy ghost zero modes make the correlator vanish,
thus cutting the instanton sum to $c=-j,\ldots,+j$.
\item
Integration over bosonic moduli cover the metric moduli (of dimension
$3g{-}3{+}n$) as well as the gauge moduli (of dimension $g{-}1{+}n$).
The latter are given by the gauge field holonomies $\oint A$ and contain
the spectral flow of the vertex operators at the punctures~\cite{KL}.
\item
$\AZI$ denotes antighost zero mode insertions, which arise from proper 
gauge-fixing in the presence of moduli; 
their precise form is not relevant here.
\item
$\PCO_\pm$ are two {\it picture-changing operators\/}, originating from susy
antighost insertions plus susy moduli integration. It is important to note 
that $\PCO_\pm$ are {\it not invertible\/} as local operators~\cite{BKL}, 
in contrast to the case of the $N{=}1$ NSR string.
\item
$\VEV{\ldots}_{g,c}(m)$ signifies a correlator of the (matter plus ghost)
conformal field theory on a worldsheet with fixed moduli $m$ and
topology~$(g,c)$.
\end{itemize}

It remains to characterize the physical string states, whose vertex
operators $V(k)$ appear in \gl{amp}.
As for the $N{=}1$ NSR string, the (open-string) BRST cohomology displays
the phenomenon of picture degeneracy, here parametrized by a pair 
$(\p_+,\p_-)$ of picture charges. 
In the picture $(-1,-1)$, the BRST analysis is easy~\cite{bien}
and leads (modulo the usual ghost zero mode doubling) 
to a single massless physical state~\cite{OVold}
\beq
\ket{-1,-1;k}\ =\ V(k)\ket{0,0;0}\qquad{\rm with}\quad\h_{\m\n} k^\m k^\n =0
\eeq
corresponding to a massless spacetime field $\f(x)$.
We take the vertex operators in \gl{amp} from this ``canonical'' picture.
The study of other pictures is simplified by the properties of the
picture-changing operators $\PCO_\pm$ 
and a spectral-flow operator~$\SFO(\a)$, $\a\in\bR$,
which commute with one another (on the cohomology) and with the BRST operator.
They shift the picture numbers as follows,
\beql
\PCO_+\ket{\p_+,\p_-}\ =\ \ket{\p_+{+}1,\p_-} \quad&,&\quad
\PCO_-\ket{\p_+,\p_-}\ =\ \ket{\p_+,\p_-{+}1} \quad, \zeile
\SFO(\a)\ket{\p_+,\p_-}\ &=&\ \ket{\p_+{+}\a,\p_-{-}\a} \quad,
\eeql
but do not lead to a commutative triangle in the $(\p_+,\p_-)$ plane 
since~\cite{BL1,BL2}
\beq\label{tpco}
\tPCO^+\ \equiv\ \PCO_-\ \SFO(+\fr12) \ \neq\
\PCO_+\ \SFO(-\fr12)\ \equiv\ \tPCO^- \quad.
\eeq
The reason is that $\SFO(\a)$ carries global boost charge $q=\a$
with respect to $L_{+-}$ whereas $\PCO_\pm$ are neutral.
Therefore, the {\it compound\/} picture-changing operators $\tPCO^\pm$
change $(\p_+,\p_-,q)$ by $(+\frac12,+\frac12,\pm\frac12)$ 
and may be employed to reach from the canonical $(-1,-1,0)$ 
all diagonal higher pictures, i.e. 
\beq
\p\ \equiv\ \p_++\p_-\ =\ -2,-1,0,+1,\ldots \qquad{\rm and}\qquad
\D\ \equiv\ \p_+-\p_-\ =\ 0 \quad.
\eeq
Subsequent application of $\SFO(\a)$ keeps~$\p$ but shifts $\D\to2\a$.
Although it has not been proved,
we assume that all physical states are obtained in this way.

Unexpectedly, we have arrived at a proliferation of physical states
at higher pictures. Distributing the indices $\pm$ in the (symmetric)
product of $2j$ compound picture-changing operators, one gets a
$(2j{+}1)$-plet of states with $q{-}\frac12\D=-j,\ldots,+j$ 
on each line of pictures with $\p=2(j{-}1)$.
Clearly, these are spin~$j$ multiplets of the broken $SL(2,\bR)$,
and $\tPCO^\pm$ itself transforms as an $SL(2,\bR)$ doublet, mapping 
$\p\to\p{+}1$ and spin~$j$ to spin $j{+}\frac12$ states~\cite{BL1,BL2}.
In detail (suppressing Chan-Paton labels),
\beq\label{states}
\begin{array}{lrl}
\p=-2:\qquad & \ket{-1,-1;k} 
\qquad\longleftrightarrow\qquad & \f(x) \\
\p=-1:\qquad & \tPCO^a \ket{-1,-1;k} 
\qquad\longleftrightarrow\qquad & \f^a(x) \\
\p=\ \ 0:\qquad & \tPCO^{(a} \tPCO^{b)} \ket{-1,-1;k} 
\qquad\longleftrightarrow\qquad & \f^{(ab)}(x) \\
\p=+1:\qquad & \tPCO^{(a} \tPCO^b\,\tPCO^{c)} \ket{-1,-1;k} 
\qquad\longleftrightarrow\qquad & \f^{(abc)}(x) \\
\qquad\vdots & \vdots\qquad\qquad\qquad\qquad\qquad & \qquad\vdots\qquad.
\end{array}
\eeq
Moreover, since spectral flow by an {\it integral\/} $\a$ 
is just a singular gauge transformation~\cite{KL},
the vertex operator for a state with integer global boost charge~$q$
creates a gauge instanton of charge~$c{=}q$ at the puncture,
\beq
\SFO(\a{=}q) \ket{c{=}0}\ \sim\ \ket{c{=}q} \quad.
\eeq
Hence, world-sheet gauge instantons fill up the $SL(2,\bR)$ multiplets.
Our assignment of $SL(2,\bR)$ spin to individual states leads to
total picture numbers of
\beq\label{selection}
\p_{tot}\ =\ 2j_{tot}-2n\ =\ 4g-4 \qquad{\rm and}\qquad \D_{tot}\ =\ -2c
\eeq
in a correlator for topology~$(g,c)$.
This selection rule is also evident from \gl{amp}.

Apparently, we find no physical states for $\p<-2$ but a growing number
for $\p>-2$. Are they physically distinct?
At this stage, two answers seem possible.
Interpretation one~\cite{BL2} sees higher-picture states 
$\f^{(a_1a_2\ldots a_{2j})}$ 
as spacetime derivatives of the single scalar field~$\f$ 
present in the canonical $\p{=}-2$ picture.
It is supported by the fact that the states $(\tPCO^\pm)^{2j}\ket{-1,-1;k}$
are polynomials of degree $2j$ in~$k^{a\ad}$, and suggests a
worldsheet--spacetime correspondence
\beq
\tPCO_a \qquad\Longleftrightarrow\qquad \z^\ad(x)\ \partder{}{x^{a\ad}}
\eeq
with some commuting $SL(2,\bR)'$ spinor~$\z(x)$.
In this way, all pictures yield the same physics, provided $\z(x)$ does not
represent a new degree of freedom.
It is tempting to relate $\z^\ad$ with extra harmonic degrees of freedom,
as present in harmonic superspace~\cite{harmonic},
and interpret picture-changing as some kind of supersymmetry. 
Interpretation two confides that there is indeed an {\it infinity\/} 
of physical states, since non-invertible picture-changing cannot identify
$\f,\f^a,\f^{(ab)},\ldots$~\cite{pope}.
However, this tower can be consistently truncated
as we shall see below.

Using the compound picture-changing operators \gl{tpco},
we condense the instanton sum in~\gl{amp},
\beq\label{amp2} 
A^{(n)}(k_1,\ldots,k_n)\ =\ \sum_{g}\ \int\! dm_h\ dm_A
\VEV{\ V(k_1)\ldots V(k_n)\ \AZI\ \ [v^a \tPCO_a]^{2j}\ }_{g,c=0}\;,
\eeq
by introducing~\cite{BL1}
\beq
\biggl( \begin{array}{c} v^+ \\ v^- \end{array} \biggr) \ =\ \sqrt{e}\
\biggl( \begin{array}{c} +\cos\fr{\q}{2} \\ -\sin\fr{\q}{2} \end{array}
\biggr)\quad.
\eeq
As the notation suggests, $A^{(n)}$ is $SL(2,\bR)$ 
(and thus $SO(2,2)$) invariant, once we insist that $v^a$ transforms
as a fundamental spinor so that the combination
\beq
\hPCO\ \equiv\ \e_{ab}\ v^a\ \tPCO^b\ =\
\sqrt{e}\ \Bigl[ \cos\fr{\q}{2}\ \PCO_+\ \SFO(-\fr12)\
+\ \sin\fr{\q}{2}\ \PCO_-\ \SFO(+\fr12) \Bigr]
\eeq
is a Lorentz singlet.
This implies that we can not only rotate away the theta angle, but also 
boost the string coupling to an arbitrary positive value~\cite{parkes}!
Obviously, the gauge instanton contributions ($c{\neq}0$) in \gl{amp}
are vital for reclaiming $SO(2,2)$ Lorentz symmetry. 
In fact, the instanton creation and destruction operators $\SFO(\pm1)$
may be interpreted~\cite{BL1} as the $SL(2,\bR)$ generators~$L_{\pm\pm}$ 
needed to restore $\cl\to SO(2,2)$.

Any choice of couplings $(e,\q)\leftrightarrow v^a$ picks a particular
real structure in spacetime and breaks $SO(2,2)\to\cl$. 
Since $SL(2,\bR)=SO(2,1)$, 
the unbroken generator $v^av^b L_{ab}$ is associated 
with the {\it null\/} vector $\vec{n}=v{\times}v$ in $\bR^{2,1}$.
Consequently, the moduli space of real structures is the light cone
in $D=2{+}1$, parameterized by $(e,\q)$ where $e$ is nothing but the
``time'' component of the lightlike~$\vec{n}$.
Since the zero instanton sector of \gl{amp} is neutral under the
$L_{+-}$ boost, it corresponds to a {\it spacelike\/} $(2{+}1)$-vector 
and can therefore not be isolated for any value of $(e,\q)$.
We may, however, eliminate all but the minimal (or maximal) instanton
sector from the amplitude~\gl{amp} by taking $\q{=}0\leftrightarrow v^-{=}0$
(or $\q{=}\p\leftrightarrow v^+{=}0$) in~\gl{amp2}.
This amounts to picking a light-cone frame in spacetime, as only
$\tPCO_+\sim\z^\ad\pa_{+\ad}$ (resp. $\tPCO_-\sim\z^\ad\pa_{-\ad}$) 
will appear.

It is straightforward to compute tree-level $n$-point amplitudes 
$A_0^{(n)}$, with the instanton sum ranging from $2{-}n$ to $n{-}2$.
{}From consistency of duality with the absence of massive states,
and also from the topological description~\cite{berk1,berk2},
we know that $A_0^{(n)}=0$ for $n{>}3$ in all instanton sectors.
The tree-level three-point function comes out as~\cite{buckow}
\beql
A_0^{(3)}\ &=\ 
\biggl<\ V^A(k_1)\ V^B(k_2)\ V^C(k_3)\ \ &\hPCO^2\ \biggr>_{0,0}
\zeile
&=\ \bigg<\ V^A(k_1)\ V^B(k_2)\ V^C(k_3)\ \Big[
&+\ v^+v^+\ \PCO_+\ \PCO_+\ \SFO({-}1) 
\zeile
&&\! -2 v^+v^-\ \PCO_+\ \PCO_-
\zeile
&&+\ v^-v^-\ \PCO_-\ \PCO_-\ \SFO({+}1) \Bigr]\ \biggr>_{0,0}
\zeile
&=\ i e f^{ABC}\ \Bigl[
\cos^2\fr{\q}{2}\ k_1^-\wedge k_2^-\
+&\sin\fr{\q}{2}\cos\fr{\q}{2}\ k_1^{(+}\wedge k_2^{-)}\
+\ \sin^2\fr{\q}{2}\ k_1^+\wedge k_2^+ \Bigr]
\eeql
using the notation of \gl{bilin}.
Employing the broken Lorentz generators, this may be written as
\beq
A_0^{(3)}\ =\ e^{\q(L_{++}-L_{--})/2}\ e^{(\ln \m e) L_{+-}}\ 
A_0^{(3)}\Big|_{\q=0,e=1/\m}
\eeq
where
\beq\label{3ptlc}
A_0^{(3)}\Big|_{\q=0,e=1/\m}\ =\ i \m f^{ABC} k_1^-\wedge k_2^-
\eeq
is the result from the $c{=}-1$ sector only. 
We have scaled $e$ to a reference value~$1/\m$, 
fixing a mass scale~$\m$.

This three-point function (and the vanishing of the higher
tree-level amplitudes) may be obtained from 
Leznov's spacetime action~\cite{leznov,parkes}
\beq\label{Leznov}
S_L\ =\ \tr\int\!d^4x\ \Bigl[
\fr12\,\f\bo\f\ +\ \fr{i}{3\m}\,\f\,(\pa_+{}^\ad\f)\,(\pa_{+\ad}\f)\Bigr]
\eeq
for a Lie-algebra-valued anti-hermitian field~$\f$.
Even better does a related two-field action
\beq\label{Siegel}
S_{CS}\ =\ \tr\int\!d^4x\ \Bigl[
\tf\bo\f\ +\ \fr{i}{\m}\,\tf\,(\pa_+{}^\ad\f)\,(\pa_{+\ad}\f)\Bigr]
\eeq
proposed by Chalmers and one of the authors~\cite{CS},
because it permits us to absorb $\m$ into the fields by 
\beq\label{rescale}
\f\ \to\ \m\,\f \quad,\qquad \tf\ \to\ \fr1\m\,\tf \quad.
\eeq
Both actions describe self-dual Yang-Mills at tree-level
and in a light-cone gauge. Their one-loop amplitudes
differ by a factor of~$2$. Beyond one loop, however,
$S_{CS}$ yields vanishing amplitudes whereas $S_L$ does not.
For further comparison and the relation to maximally
helicity violating pure Yang-Mills amplitudes see~\cite{CS}.

Another light-cone gauge, proposed by Yang~\cite{yang}, 
leads to a {\it non-polynomial\/} effective action~\cite{DNS}
\beq\label{Yang}
S_Y\ =\ \tr\int\!d^4x\ \Bigl[
\fr12\,\f\bo\f\ +\ \fr{i}{3\m}\,\f\,(\pa_{(+}{}^\ad\f)\,(\pa_{-)\ad}\f)\
+\ O(\f^4) \Bigr]
\eeq
which was shown to reproduce the zero-instanton ($c{=}0$) sector of 
the $N{=}2$ open string tree-level amplitudes~\cite{OVold,marcus}.
Yang's action \gl{Yang} also has a two-field relative~\cite{CS}.
A point we make in this paper is that the full string amplitude
necessarily receives gauge instanton contributions, which
shift the spacetime effective action from the Yang to the
Leznov type. 
Not only is the latter {\it polynomial\/} (cubic) but it also allows
us to parametrize the explicit breaking of $SO(2,2)$ Lorentz invariance
with a {\it spacetime twistor\/} made from the coupling constants $(e,\q)$.

At tree level, the closed string is basically the square of the
open string, so our considerations apply just as well to the closed
$N{=}2$ string. Here, the instanton contributions are labelled by
a pair $(c_L,c_R)$, 
and are summed over independently for left- and right-movers~\cite{OVnew}.
The zero-instanton sector is known~\cite{OVold} to be described
by the (cubic) Pleba\'nski action~\cite{ple}
\beq\label{Ple}
S_P\ =\ \int\!d^4x\ \Bigl[ \fr12\,\vf\bo\vf\ +\ \fr{2}{3\m^3}\,\vf\,
(\pa_+{}^\ad \pa_-{}^\bd \vf)\,(\pa_{+\ad} \pa_{-\bd} \vf) \Bigr]\quad,
\eeq
but the instanton sum again corrects this to a Leznov-type action,
\beq
S'_L\ =\ \int\!d^4x\ \Bigl[ \fr12\,\vf\bo\vf\ +\ \fr{2}{3\m^3}\,\vf\,
(\pa_+{}^\ad \pa_+{}^\bd \vf)\,(\pa_{+\ad} \pa_{+\bd} \vf) \Bigr]
\eeq
or its two-field cousin using a Lagrange multiplier field~\cite{CS}. Here, 
$\vf$ denotes the K\"ahler deformation of self-dual gravity~\cite{OVold}.

It is possible to obtain the two-field action~\gl{Siegel} 
(after the rescaling~\gl{rescale}) by an appropriate assignment
of spacetime fields to string vertex operators. 
Let us indicate $\p$ charges by braced subscripts, i.e.~$V_{(\p)}$.
Note that only the highest component of each $SL(2,\bR)$ multiplet
appears, as picked out by the twistor~$v^a$.
Since $\f$ and $\tf$ carry $GL(1,\bR)$ charges $q{=}-2$ and $q{=}+2$,
respectively, and the number of Lagrange multiplier legs $\tf$ in
nonvanishing amplitudes is $1{-}g$ while the total picture
increases by 4 units per loop, we are forced to associate
\beq\label{identify}
\f\ \longleftrightarrow\ V_{(0)} \qquad{\rm and}\qquad
\tf\ \longleftrightarrow\ V_{(-4)} \quad.
\eeq
This seems problematic because
there are no physical states in the picture $\p=-4$.
Yet, it is possible to define such states by creating a pairing
\beq
\VEV{\ V_{(-4-\p)}(-k)\ V_{(\p)}(k)\ }_0\ =\ 1
\eeq
via the reflection symmetry (nonlocal on the worldsheet)
\beq
\p\ \leftrightarrow\ -4-\p \quad,\qquad
\D\ \leftrightarrow\ -\D \quad,\qquad
q\ \leftrightarrow\ -q \quad,\qquad
k\ \leftrightarrow\ -k \quad.
\eeq
The correspondence~\gl{identify} allow us to identify (spacetime)
helicity~$s$ with $GL(1,\bR)$ charge~$q$ and picture number~$\p$ 
through $s=\fr12q=1{+}\fr12\p$, and the 
maximal helicity condition~\cite{CS} is nothing but \gl{selection}.

In order to restrict the helicities to $|s|\le1$, a truncation
to $-4\le\p\le0$ is necessary.  Employing
the vanishing of all tree-level $(n{>}3)$-point functions~\cite{berk2}
and the BRST-exactness of $\pa\PCO_\pm$ and $\pa\SFO$,
\beql
0\ &=&\ \VEV{\ V^1_{(-2)}\ldots V^{n-1}_{(-2)}\ V^n_{(-2)}\
\AZI\ [\PCO_{(+1)}]^{2n-4}\ }_{g=0}
\zeile
&=&\ \VEV{\ V^1_{(-2)}\ldots V^{n-1}_{(-2)}\ V^n_{(2n-6)}\ \AZI\ }_{g=0}
\quad,
\eeql
we may set $V_{(\p>0)}$ to zero, 
keep only the pictures from $-4$ to~$0$.
At the loop level, moving all $\PCO$ onto a single vertex produces a
$V_{(4g+2n-6)}$, which would kill the correlator for $n>3{-}2g$, 
according to our truncation~$\p{\le}0$. 
As the explicit non-vanishing of the one-loop three-point 
function~\cite{bonini} shows, however, this argument is too naive, 
and we expect it to be modified by the appearance of contact terms and 
moduli boundary terms.\footnote{
Moving $\PCO$ on the worldsheet produces extra correlators containing
$\{Q_{\rm BRST},\AZI\}$.}
A further restriction to the extremal helicities $s{=}\pm1$
(i.e. $\p=-4,0$ only) yields the two-field action~\gl{Siegel}.

It is instructive to work out the 4D mass dimensions of 
worldsheet objects. After rescaling~\gl{rescale}, we learn from
\gl{Siegel} that $[\f]=0$ and $[\tf]=2$.
Since the vertex operators multiply the spacetime fields in
the worldsheet action, it follows that $[V_{(0)}]=0$ and
$[V_{(-4)}]=-2$ which means that $[\PCO]=+\fr12$,
as is appropriate for spacetime supersymmetry.

It may well be that the $N{=}2$ string contains more than just
self-dual Yang-Mills or gravity. If the semi-infinite picture tower of
massless physical states is for real, its proper spacetime identification
presumably requires to consider more general than self-dual backgrounds.

\bigskip\noindent
O.L. acknowledges discussions with G. Chalmers, E. Kiritsis and C. Kounnas.



\end{document}